\documentclass[twocolumn,showpacs,fleqn,nobibnotes]{revtex4}
\usepackage{amsmath}
\usepackage{graphicx}
\usepackage{float}
\usepackage{subfigure}
\usepackage{verbatim}
\usepackage{color}
\usepackage{comment}
\usepackage{hyperref}

\begin{document}
	
	\title{Coherent and incoherent Upsilon production in ultraperipheral collisions at the Large Hadron Collider}
	\pacs{12.38.Bx; 13.60.Hb}
	\author{M.B. Gay Ducati, F. Kopp, M.V.T. Machado}
	
	\affiliation{High Energy Physics Phenomenology Group, GFPAE  IF-UFRGS \\
		Caixa Postal 15051, CEP 91501-970, Porto Alegre, RS, Brazil}

\begin{abstract}

The exclusive photoproduction of $\Upsilon (nS)$ states were calculated in ultra-peripheral collisions for coherent and incoherent process in PbPb at $\sqrt{s_{NN}}$= 5.5 TeV. Different dipole models were compared in the theoretical framework of light-cone color dipole formalism. Moreover, it was calculated the differential cross section for the Upsilon states and their total cross section for two intervals of rapidity: $|y|\leq 4$ 4 and $ 2 \le$ y $\le 4.5$. A systematic study is done on the theoretical uncertainties associated to the production and predictions are presented for the first time for the incoherent cross section of the radially excited states. 

\end{abstract}	

\maketitle
	
\section{Introduction}	

The study of exclusive meson photoproduction in UPC \cite{upcs} is an essential tool to understand the low-x physics and also to investigate the gluon density in this regime. In the UPC case, the exclusive photoproduction dominates the process through the emission of quasi-virtual photons which interact with the target. The photon-target interaction amplitude, when considering the light-cone dipole formalism \cite{nik}, can be written as a convolution between the photon-meson wave functions overlap and the elementary dipole-target cross section \cite{Nemchik:1996pp}. In addition, the process considered here is quasi-elastic, $Q^{2}$ $\approx$ 0, and in the region of small-x, the gluon density may increase to the point where gluon fusion, gg $\to$ g, becomes significant. This kind of fusion produces nonlinear effects in the evolution equations. For instance, at mid-rapidity the typical value of Bjorken variable is $x = \frac{m_{\Upsilon}}{\sqrt{s_{AA}}}\simeq 10^{-3}$ for PbPb collisions at the LHC. Differently from DGLAP equations which is a linear equation, dipole models incorporate linear and non-linear effects\cite{CGC}. In pQCD the exclusive meson photoproduction has a differential cross-section $\propto$ [$xg(x,Q^{2})]^{2}$, where the gluon distribution functions do not take into account the effects of saturation since they are evolved by DGLAP equations. The inclusion of parton saturation and nuclear effects are crucial in describing even the experimental observations in heavy-ion collisions at RHIC (for instance, see Ref. \cite{heavy}). Within the color dipole approach one can introduce information on dynamics beyond the leading
logarithmic QCD approach for meson production and computing predictions for the radially excited states is a reasonably easy task \cite{Nemchik:1996pp}. From the experimental point of view, the considered process is quite clear due to the presence of two rapidity gaps. The rapidity gap describes a region between the axis beam and the decay of meson $(e^{+}+e^{-},\mu^{-} +\mu^{+})$ free of particles. In this way, the experimental observation of this kind of quasi-diffractive process is facilitated. In addition, the absorption corrections in this case is not strong as in the corresponding final state in proton-proton collisions.

In this work, we investigate the exclusive (coherent and incoherent) production of
$\Upsilon (1S)$ and its radially excited states $\Upsilon (2S)$ and $\Upsilon (3S)$
in PbPb collisions for LHC energy. In a previous work \cite{GKMM} the coherent photoproduction of $\Upsilon$ states at various energies in $pp$, pPb PbPb collisions at the LHC has been considered. Those calculations were carried out in the theoretical framework of the color light-cone dipole formalism \cite{nik} and focused only on the coherent channel where the initial state particles remain intact after interaction. It was shown that the corresponding predictions describe correctly the experimental results from LHCb Collaboration \cite{UpsilonLHCb} for $\Upsilon$ photoproduction in $pp$ collisions. Those data were obtained for typically large rapidities and the $x$-values to be covered are increasingly smaller for forward rapidities. We roughly get $x=\frac{m_{\Upsilon}}{\sqrt{s_{AA}}}e^{-y}\simeq 8\times 10^{-5}$ at $y=3$ and it is clear that gluon dynamics is being probed at extremely low-$x$ and low perturbative scales $\mu^2 \simeq 20$ GeV$^2$. This kinematical range is in the limit of application of usual pQCD and saturation approach should be relevant.  For nuclear targets, the nuclear saturation scale $Q_{sat,A}^2\simeq c A^{1/3}Q_{sat,p}^2$ (with $c\simeq 0.3$) reaches $2$ GeV$^2$ in those cases \cite{Dusling}. The main novelty in current work is the detailed study of the incoherent cross section for the upsilon states. This is quite important, as it was pointed out that incoherent diffraction probes the fluctuations in the interactions strengths of multiparton Fock states in the nuclear wavefunctions \cite{Lappi}. The connection between incoherent diffraction and fluctuations is a quite rich subject and the pioneering works are found in Refs. \cite{FlucHis}. Recently, the topic is very  active and we call attention to the following works \cite{recworks,gbwnew}.

The paper is organized as follows. In the next section we give the main theoretical information to obtain the rapidity distribution of coherent and incoherent production of $\Upsilon (1S,2S,3S)$ states in PbPb collisions for the future LHC run and energies close to the run2. The main motivation is the successful description of experimental results  measured by LHCb Collaboration \cite{UpsilonLHCb} for $\Upsilon (1S)$ in $pp$ collisions. In the section \ref{discussions} we present the phenomenological calculations, discuss the main theoretical uncertainties and a comparison with other approaches is performed. For instance, we compare the present calculation to the predictions available using the STARlight Monte Carlo \cite{STARlight,KN}. Finally, we show the main conclusions.

\section{Theoretical Framework}

 The exclusive meson photoproduction in nucleus-nucleus collisions can be factorized in terms of the equivalent flux of photons of the nucleus projectile and photon-target production cross section \cite{upcs}. In UPCs there is the absence of strong interactions between the projectile particle and the target. In this case the reaction is characterized by impact parameter $>$ 2 $R_{A}$ and as the interaction is ultra-relativistic and purely electromagnetic, one can use the Weizs\"{a}cker-Williams approximation \cite{upcs}. The photon energy spectrum, $dN_{\gamma}^A/d\omega$, which depends on the photon energy $\omega$,  is well known \cite{upcs}. The rapidity distribution for $\Upsilon$ states  photoproduction in  $AA$ collisions can be  written down as,
\begin{eqnarray}
\frac{d\sigma}{dy} (A A \rightarrow   A\otimes \Upsilon(nS) \otimes Y) & = & \left[ \omega \frac{dN_{\gamma}^A}{d\omega }\,\sigma(\gamma A \rightarrow \Upsilon(nS)\,Y ) \right. \nonumber \\
& + & \left. \left(y\rightarrow -y \right) \right],
\label{dsigdyA}
\end{eqnarray}
where the photon flux in nucleus is denoted by $dN_{\gamma}^A/d\omega$ and $Y=A$ (coherent case) or $Y=A^*$ (incoherent case). The symbol $\otimes$ denotes the large rapidity gap between the produced meson and the final states nucleus.

 The produced state with mass $m_V$ has rapidity $y\simeq \ln (2\omega/m_V)$ and the square of the $\gamma A$ centre-of-mass energy is given by $W_{\gamma A}^2\simeq 2\omega\sqrt{s}$.  The photon-Pomeron interaction will be described within the light-cone dipole frame, where the probing
projectile fluctuates into a
quark-antiquark pair with transverse separation
$r$ (and momentum fraction $z$) long after the interaction, which then
scatters off the hadron. The cross section for exclusive photoproduction of $\Upsilon$ states  off a nucleon target is given by,
\begin{eqnarray}
\sigma (\gamma p\rightarrow \Upsilon \,p) = \frac{\left|\sum_{h, \bar{h}} \int dz\, d^2r \,\Psi^\gamma_{h, \bar{h}}\sigma_{dip}(x,r)\, \Psi^{V*}_{h, \bar{h}}  \right|^2}{16\pi B_{\Upsilon}} ,
\label{sigmatot}
\end{eqnarray}
where $\Psi^{\gamma}$ and $\Psi^{V}$ are the light-cone wavefunction  of the photon  and of the  vector meson ($V = \Upsilon $), respectively.  The dipole-proton cross section is denoted by  $\sigma_{dip}(x,r)$ and the  diffractive slope parameter by $B_V$.  In this context, we are implicitly assuming that the
proton shape is Gaussian and that the impact parameter dependence
factorizes out from the dipole-nucleon scattering amplitude. Here, we consider the energy dependence of the slope using the Regge motivated expression \cite{GKMM}.  

 The exclusive photoproduction off nuclei for coherent and incoherent processes can be simply computed in high energies where the large coherence length $l_c\gg R_A$ is fairly valid. The expressions for both cases are given by \cite{Boris},
\begin{eqnarray}
\sigma (\gamma A \rightarrow \Upsilon A) & = & \int d^2b\, \left| \sum_{h, \bar{h}} \int dz\, d^2r \,\Psi^\gamma_{h, \bar{h}} \Psi^{V*}_{h, \bar{h}} \right. \nonumber \\
&\times & \left. \left[1-\exp\left(-\frac{1}{2}\sigma_{dip}(x,r) T_A(b)\right)  \right]\right|^2, \label{eq:coher} \nonumber \\
\sigma (\gamma A \rightarrow \Upsilon A^* )  & = & \int d^2b\,\frac{T_A(b)}{16\pi\,B_V}\left| \sum_{h, \bar{h}} \int dz\, d^2r \,\Psi^\gamma_{h, \bar{h}} \Psi^{V*}_{h, \bar{h}} \right. \nonumber \\
 &\times & \left. \sigma_{dip}(x,r) \exp\left[-\frac{1}{2}\sigma_{dip}(x,r)T_A(b)  \right]\right|^2, \nonumber
\label{eq:incoh}
\end{eqnarray} 
where $T_A(b)= \int dz\rho_A(b,z)$  is the nuclear thickness function. In the numerical evaluations, we have considered the boosted Gaussian wavefunction and several phenomenological saturation models, which encode the main properties of the saturation approaches.  Accordingly, the cross sections above include both the skwedness and real part of amplitude corrections. Namely, we multiply the result above by $K^2 = R_g^2 (1 + \beta^2)$, where $\beta = \tan (\pi \lambda_{ef}/2)$ is the ratio of real to imaginary parts of the scattering amplitude and $R_g $ incorporates the off-forward correction (see \cite{GKMM} for details). The effective power on energy, $\lambda_{ef}$ is determined for each case. In order to take into account the threshold correction for the dipole cross section, we have multiplied them by a factor $(1-x)^7$.

Finally, we set the parameters and phenomenological models to be considered in next section. For the slope parameter it was considered the energy dependency from the Regge phenomenology \cite{GKMM},
\begin{eqnarray}
B_{\Upsilon}=b^{\Upsilon}_{el}+2\alpha'\text{log}\left(\frac{W^2_{\gamma A}}{W^2_0}\right) 
\end{eqnarray}
with $\alpha'=0.164$ GeV$^{-2}$, $W_{0}=95$ GeV, $b_{el}^{\Upsilon_{(1S)}}=3.68$ GeV$^{-2}$, $b_{el}^{\Upsilon_{(2S)}}=3.61$ GeV$^{-2}$ and $b_{el}^{\Upsilon_{(3S)}}=3.57$ GeV$^{-2}$. It will be taken into account only for the incoherent cross section. For the meson wavefuntion, we will use the Boosted-Gaussian model \cite{wfbg} because it can be applied in a systematic way for excited states. The corresponding function is given by \cite{Sanda2},
\begin{eqnarray}
\phi_{nS}(r,z) = \left[\sum^{n-1}_{k=0}\alpha_{nS,k}R^2_{nS}\hat{D}^{2k}(r,z)\right]G_{nS}(r,z),
\end{eqnarray}
with $\alpha_{nS,0}=1$. The operator $\hat{D}^{2}(r,z)$ is defined by
\begin{eqnarray}
\hat{D}^{2}(r,z) = \frac{m_f^2-(\frac{1}{r}\partial_r+\partial^2_r)}{4z(1-z)}-m_f^2,
\end{eqnarray}
 and it acts on the following generatrix function 
\begin{eqnarray}
 G_{nS}(r,z) & = & \mathcal{N}_{nS}\,z(1-z)\,\exp\left(-\frac{m^2_f\mathcal{R}^2_{nS}}{8z(1-z)} \right. \nonumber \\
& - & \left. \frac{2z(1-z)r^2}{\mathcal{R}^2_{nS}}+\frac{m^2_f\mathcal{R}^2_{nS}}{2}\right).
\end{eqnarray}

The main physical quantity is the dipole scattering cross section. We consider the following phenomenological models in our analysis: GBW \cite{PRD59-014017}, CGC \cite{PLB590-199} and BCGC \cite{PRD74-074016}. The GBW model is defined by the eikonal shape for the dipole cross section,
\begin{eqnarray}
\sigma^{GBW}_{q\bar q}(x,r)=\sigma_0\left(1-e^{-r^2Q^2_s(x)/4}\right)\label{gbw},
\end{eqnarray}
where $\sigma_0=2\pi R^2$ is a constant and $Q_s^2(x)=(x_0/x)^{\lambda}$ GeV$^2$ denotes the saturation scale.  We also consider the CGC model \cite{PLB590-199}, based in the Color Glass Condensate framework, in which gluon saturation effects are incorporated via an approximate solution of the Balitsky-Kovchegov equation \cite{CGC}.  The expression for the CGC model is given by,
\begin{eqnarray*}
\sigma^{CGC}_{q\bar q}(x,r) = \sigma_0\begin{cases}
\mathcal{N}_0\left(\frac{rQ_s}{2}\right)^{\gamma_{eff}(x,r)} & :rQ_s\leq 2
\\
1-e^{-Aln^2(BrQ_s)} & :rQ_s>2
\end{cases}
\label{iim}
\end{eqnarray*}
where $\gamma_{eff}(x,r)=2\left(\gamma_s+(1/\kappa\lambda \,\ln (1/x))ln(2/rQ_s)\right)$ is the effective anomalous dimension and one has the constant $\kappa=9.9$.  

In order to investigate the theoretical uncertainty associated to the models for the dipole cross section, we use the original values (OLD label) of parameters for the fits  including the charm contribution. That is, for GBW-OLD we follow Ref. \cite{PRD59-014017}, for CGC-OLD Ref. \cite{Soyez} is considered and bCGC-OLD refers to Ref. \cite{PRD74-074016}. The bCGC model uses the same functional form of Eq. (\ref{iim}) and replaces the saturation scale in the following way: $Q_s^2(x) \rightarrow Q_s(x,b)^2= (x_0/x)^{\lambda}\exp[-b^2/(2\gamma_sB_{CGC})]$. Moreover, we consider the updated version of those models, GBW-NEW \cite{gbwnew}, CGC-NEW \cite{Amir} and bCGC-NEW \cite{Amir}, respectively. A comment is in order here: the GBW-NEW parametrization  is very different from other color dipole fits, as it includes energy evolution of the
subnucleonic shape of the proton and it can
potentially significantly affect the incoherent cross section. In particular, GBW-OLD and GBW-NEW are qualitatively very
different and GBW-NEW was not fitted to all $F_2$ small-$x$ data (DESY-HERA) as discussed in details in Ref. \cite{gbwnew}.

\section{Results and discussions}
\label{discussions}

\begin{figure*}[t]
	\begin{tabular}{ccc}	
	\includegraphics[scale=0.4]{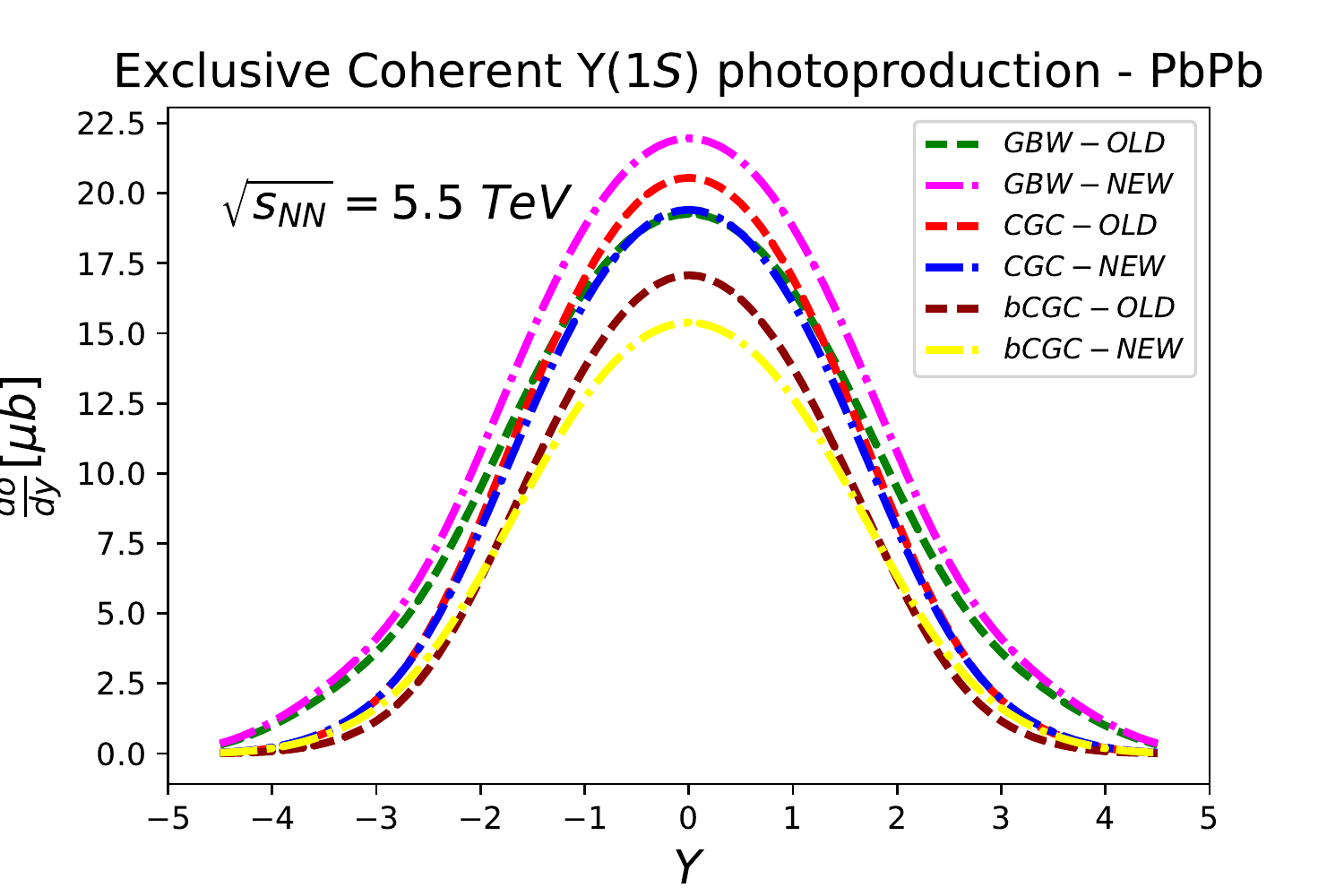} &
	\includegraphics[scale=0.4]{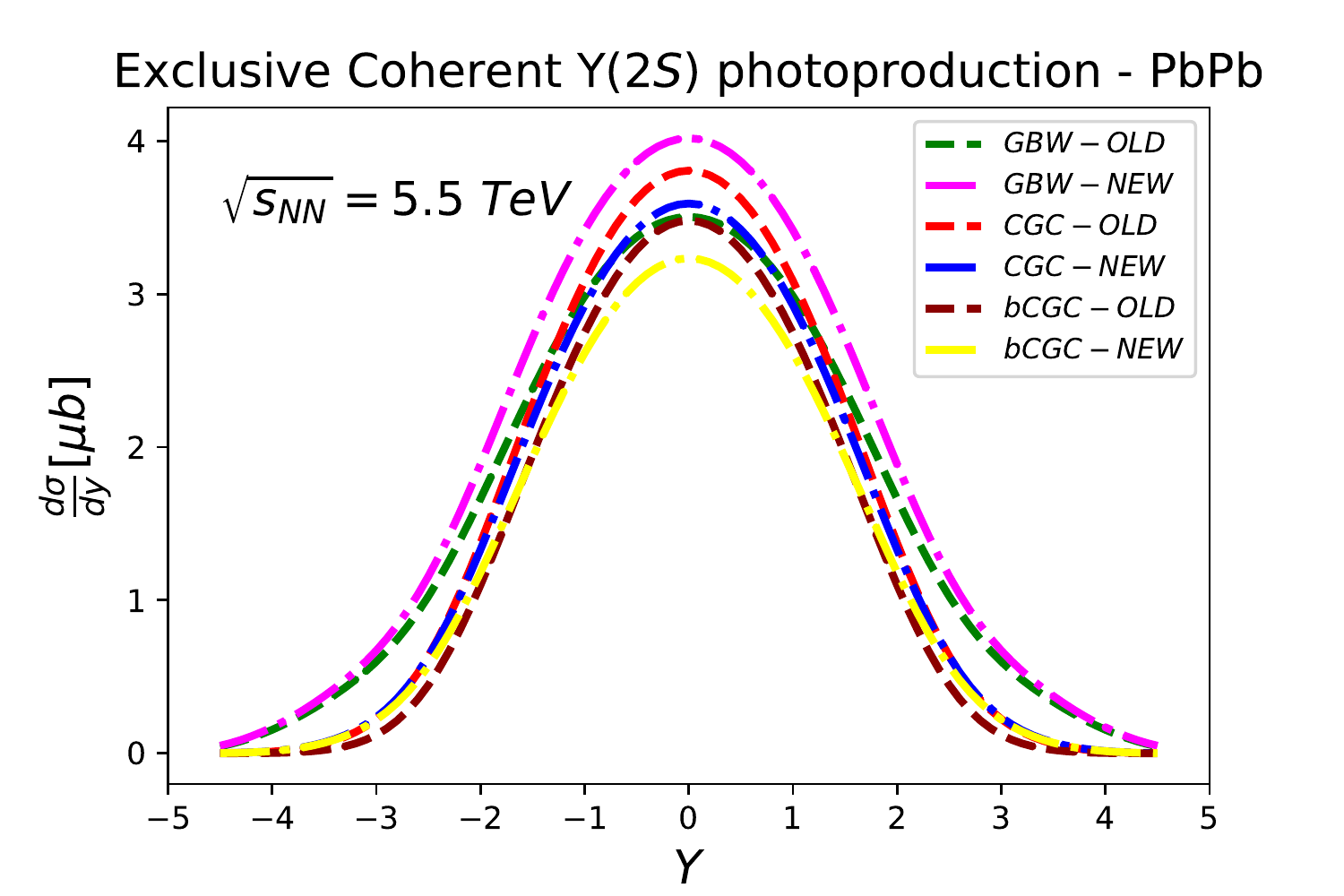} &   
	\includegraphics[scale=0.4]{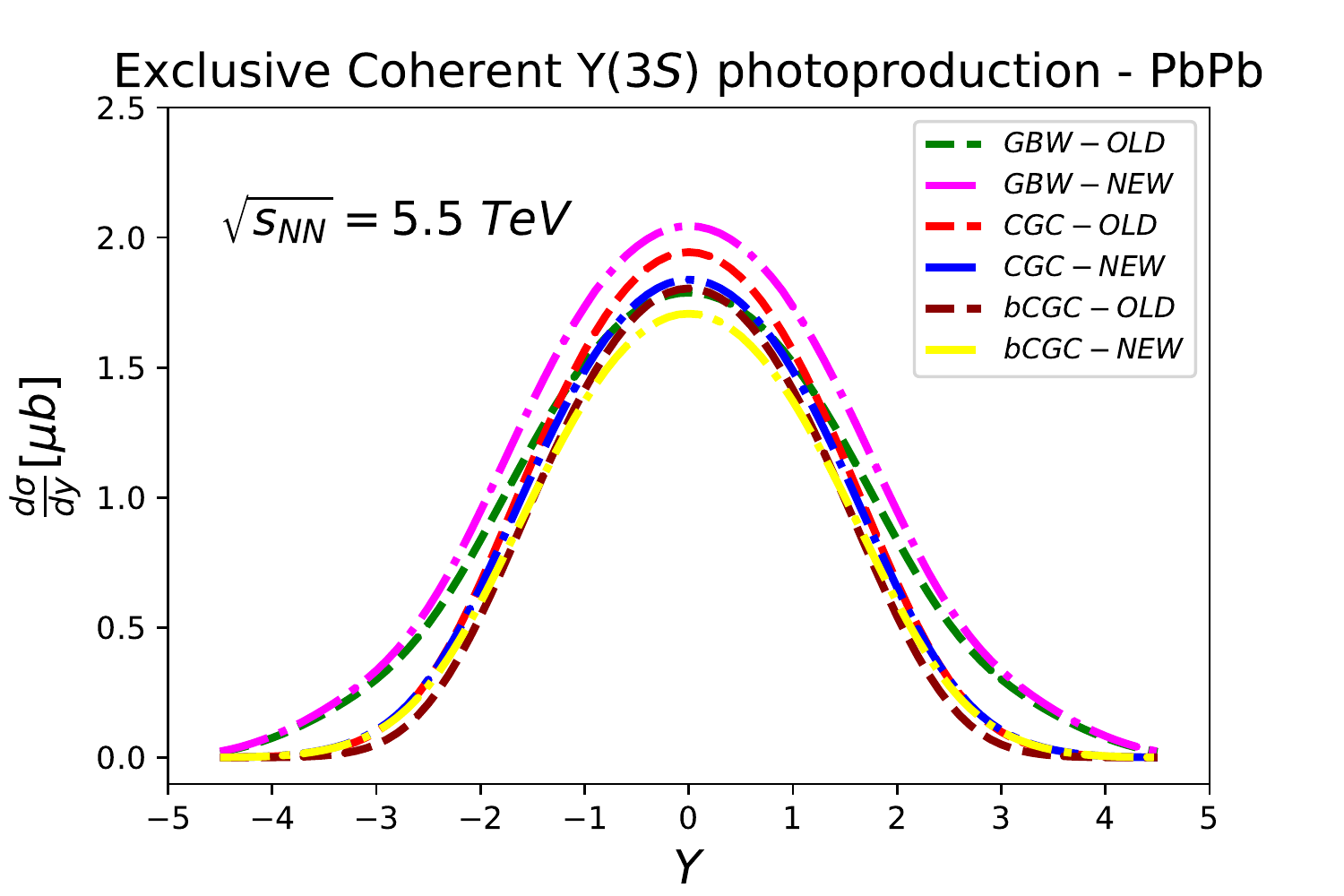}   
		
		\end{tabular}
		\caption{Exclusive coherent photoproduction of $\Upsilon(1S,2S,3S)$ in PbPb at $\sqrt{s} = 5.5$ TeV for GBW, CGC and bCGC dipole models.}
		\label{fig:1}
	\end{figure*}

\begin{figure}[t]
			\includegraphics[scale=0.5]{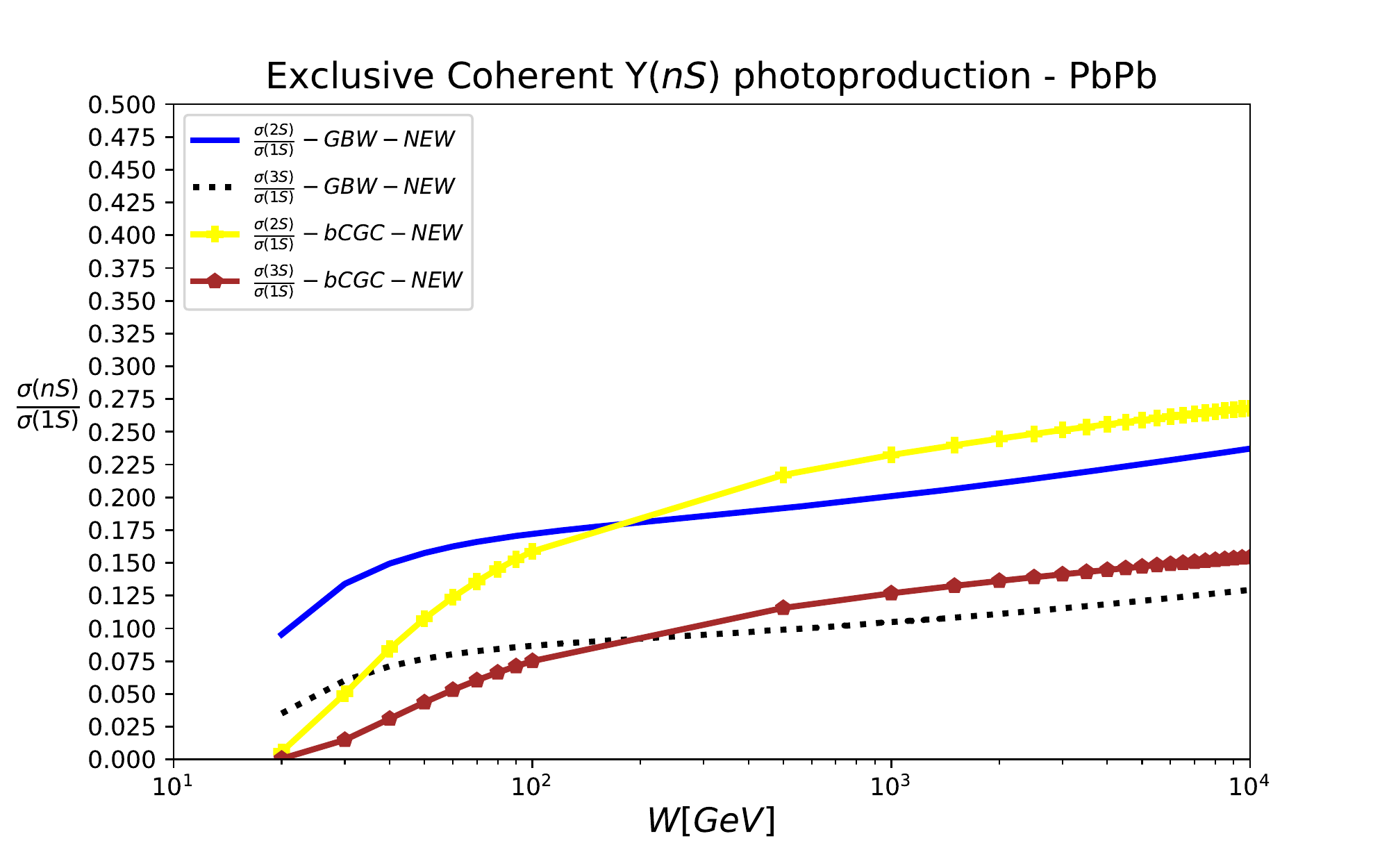}
						\caption{ The ratio of the cross sections, $\sigma(\gamma A \rightarrow \Upsilon (nS)A)/ \sigma(\gamma A \rightarrow \Upsilon (1S)A)$, as a function of $W_{\gamma A}$ for the parametrizations bCGC-NEW and GBW-NEW. }
		\label{fig:2.2}
	\end{figure}

Let us start the analysis by computing the theoretical predictions for the coherent process for PbPb collisions at 5.5 TeV. Here, we disregard any absorptive corrections. In Figure $\ref{fig:1}$, it is presented the results for photoproduction of $\Upsilon$ states, including its radial excitations, taking into account the different models presented in the last section. The theoretical uncertainty is relatively large, being of order 15 \% for the $1S$ state (similar for the remaining $2S$ and $3S$ states). We could have an additional uncertainty related to the vector meson wave function, however in Ref. \cite{GSSM2} it was shown that this is not the case for $\Upsilon$ states (the overall theoretical uncertainty is within the experimental error bars in $pp$ case \cite{UpsilonLHCb}).  For the main contribution, we have $d\sigma_{coh} /dy\, (y=0)= 18.5\pm 3.5$ $\mu$b for $\Upsilon (1S)$. The relative contribution of the excited states compared to the bound states is $\Upsilon(1S)/\Upsilon(2S)/\Upsilon(3S)=1/0.17/0.09$. We see that the relative normalization and the overall behavior is changed mostly at mid-rapidity when comparing the old and updated versions of the dipole cross sections (the deviation at large rapidities is less evident). Notice that the LHCb data for upsilon production in $pp$ collisions is reproduced by all the models in the forward region \cite{UpsilonLHCb} as shown in Ref. \cite{GKMM}. Therefore,the current level of the experimental uncertainties does not allow us to make definitive statements about the precision of the models considered. For sake of completeness, we present the integrated cross sections considering distinct cuts on rapidity. In Table \ref{TabI}, we present the results for the full rapidity coverage, $-4<y<4$, and forward rapidities, $2\leq y\leq 4.5$. In both tables \ref{TabI} and \ref{TabII}, we present only the updated versions of the dipole cross sections. For sake of completeness, we present the ratio of the cross sections, $\sigma(\gamma A \rightarrow \Upsilon (nS)A)/ \sigma(\gamma A \rightarrow \Upsilon (1S)A)$, as a function of the photon-nucleus centre-of-mass energy, $W_{\gamma A}$. We present in Fig. \ref{fig:2.2}  the result using the bCGC-NEW and GBW-NEW dipole cross sections. It was verified that the CGC-NEW result is quite similar to the bCGC-NEW one. We see a relative energy dependence, following the same trend as for the $\psi(nS)$ states \cite{armesto}. 
	
	\begin{figure*}[t]
		\begin{tabular}{ccc}
			\includegraphics[scale=0.4]{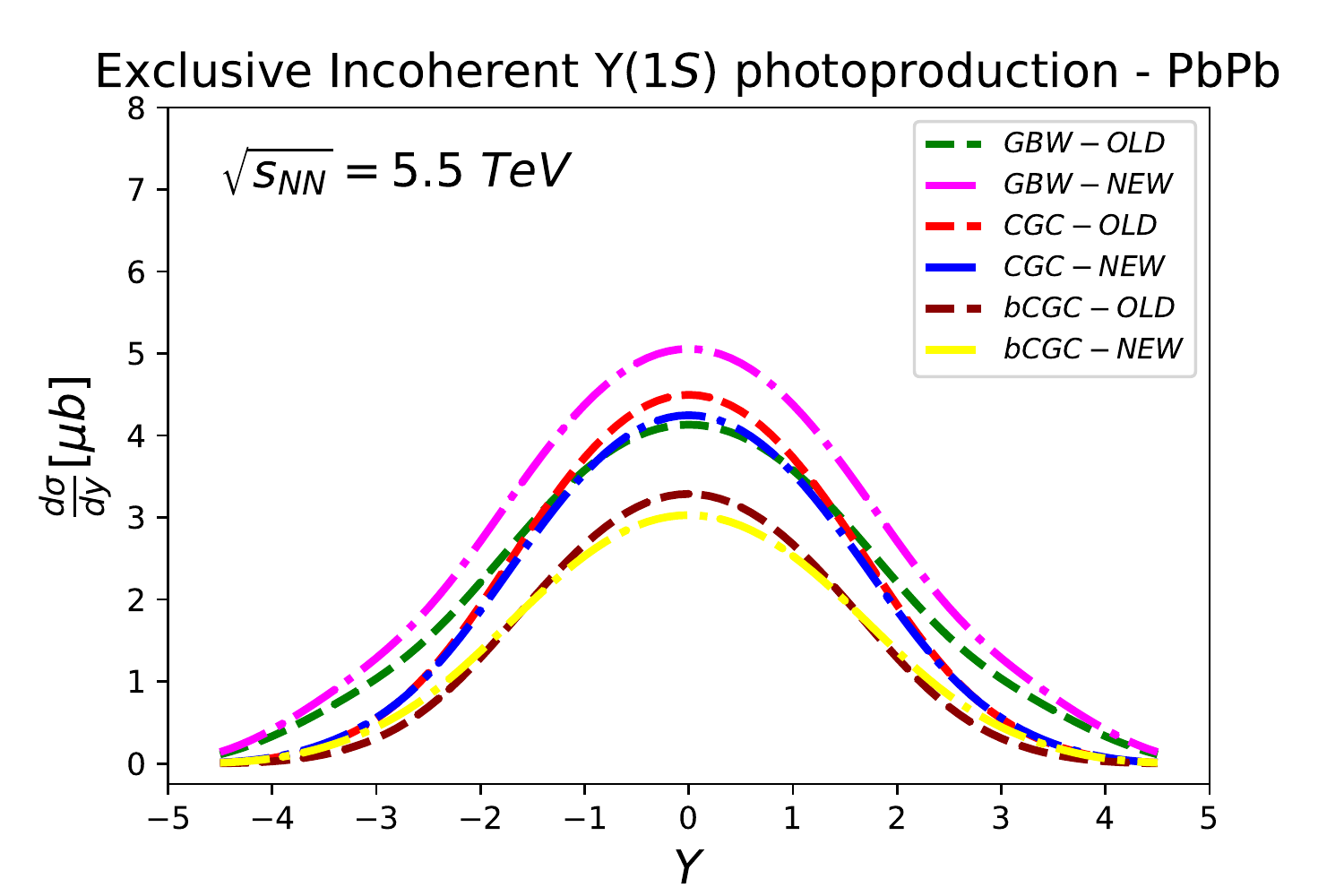}&
			\includegraphics[scale=0.4]{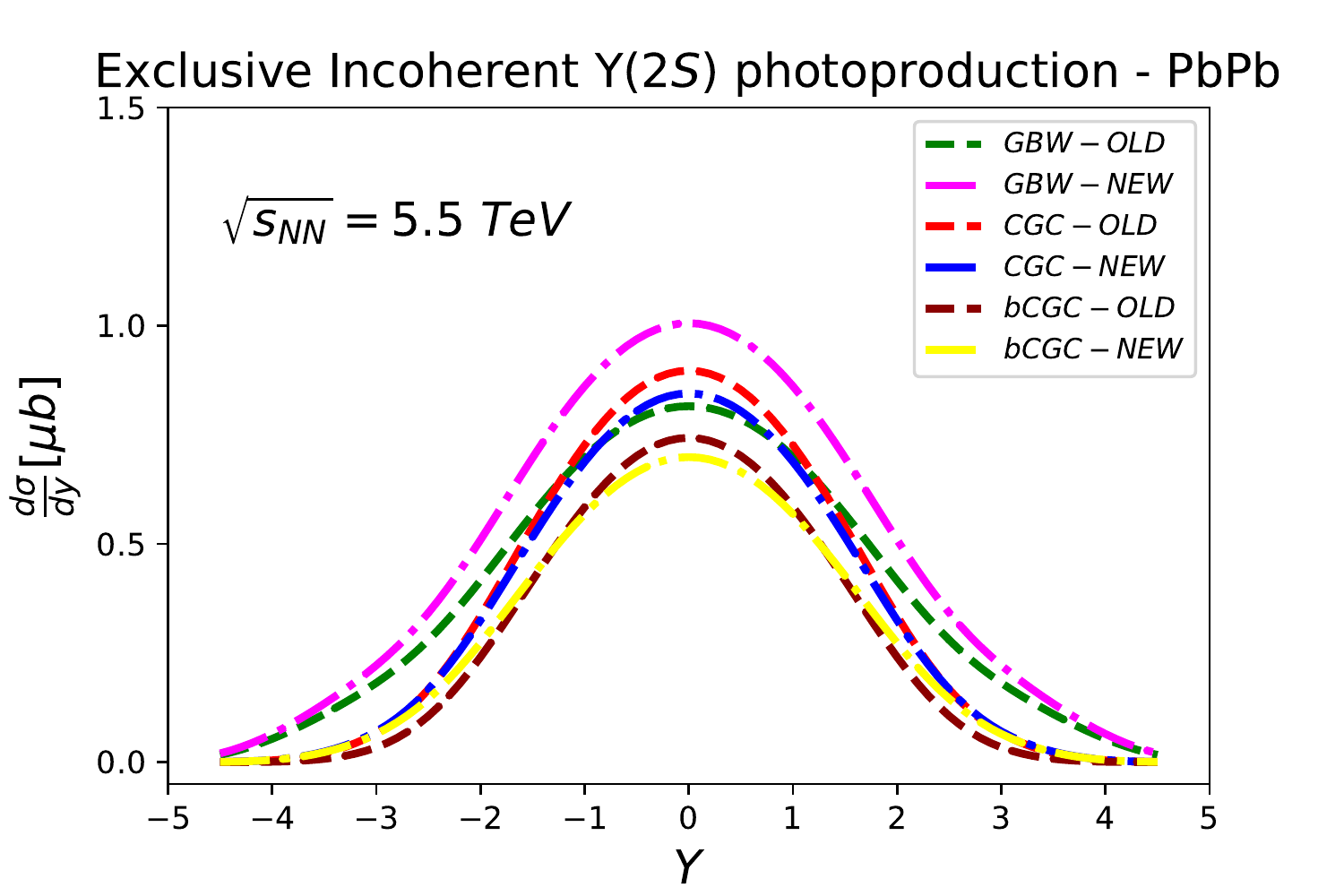} &
			\includegraphics[scale=0.4]{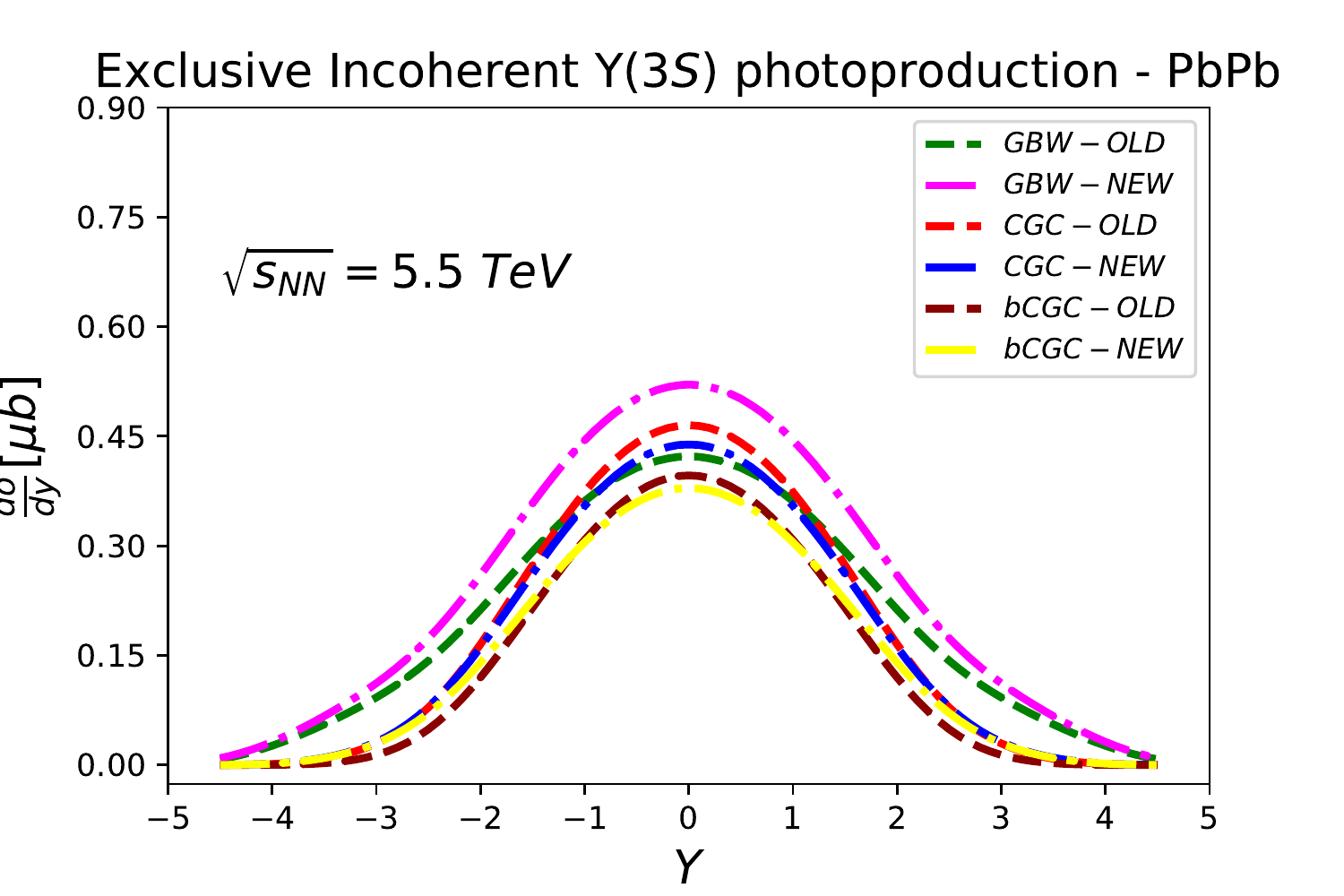}

	    \end{tabular}
		\caption{Exclusive incoherent photoproduction of $\Upsilon(1S,2S,3S)$ in PbPb at $\sqrt{s} = 5.5$ TeV for GBW, CGC and bCGC dipole models. }
		\label{fig:2}
	\end{figure*}

We now focus on the incoherent reaction, $PbPb\rightarrow Pb\,\Upsilon\, Pb^*$. This is a new contribution to the literature concerning the upsilon production. The rapidity distribution is shown in Fig. \ref{fig:2} using the same notation as the previous figure. As already known, the incoherent cross section is smaller than the coherent one. The typical ratio is $(d\sigma_{inc}/dy)/(d\sigma_{coh}/dy)\simeq 0.2$. For instance, we obtain $d\sigma_{inc} /dy\, (y=0)= 3.75\pm 1.25$ $\mu$b for $\Upsilon (1S)$. The theoretical uncertainty seems to be larger than in the coherent case. The integrated cross sections are shown in Table \ref{TabII} in  the rapidity ranges $|y|\leq 4$ and $2\leq y\leq 4.5$.

\begin{table}[t]
	\caption{Integrated cross section (in units of $\mu$b) for coherent reactions, $PbPb\rightarrow Pb\,\Upsilon\, Pb$, for full rapidity coverage (and forward rapidities). Here, we consider the updated versions of dipole cross sections.} 
	\centering 
	\begin{tabular}{c c c c} 
		\hline\hline 
		process: PbPb & $ \sqrt{s}=5.5 $  TeV    &\ \      $ |y|\leq 4\,\,  (2\leq y\leq 4.5)$  &  \\ [0.5ex] 
		$\Upsilon(nS)$ & GBW & CGC & b-CGC  \\ [0.5ex] 
		$\Upsilon(1S)$ & 163.7 \,(60.8) & 171.9 \,(63.8)   & 143 \,(53.1)   \\ %
		$\Upsilon(2S)$ & 20.3\,(7.8) &  22.0\,(8.2)  &  20.5\,(7.7) \\
		$\Upsilon(3S)$ & 10.3 \,(3.9) & 11.9\,(4.3)  &   10.9\,(4.1) \\
		[1ex] 
		\hline 
	\end{tabular}
	\label{TabI} 
\end{table}

\begin{table}[t]
	\caption{Integrated cross section (in units of $\mu$b) for incoherent reactions, $PbPb\rightarrow Pb\,\Upsilon\, Pb^*$, for full rapidity coverage (and forward rapidities).Here, we consider the updated versions of dipole cross sections.} 
	\centering 
	\begin{tabular}{c c c c} 
		\hline\hline 
		process: PbPb & $ \sqrt{s}=5.5 $  TeV    &\ \      $|y|\leq 4\,\,  (2\leq y\leq 4.5)$  &  \\ [0.5ex] 
		$\Upsilon(nS)$ & GBW & CGC & b-CGC  \\ [0.5ex] 
		$\Upsilon(1S)$ & 61.2\,(25.6) & 58.5\,(24.4)  &   44.5\,(18.6) \\ %
		$\Upsilon(2S)$ & 8.2\,(3.4) &  8.0\,(3.4)  &   6.9\,(2.9) \\
		$\Upsilon(3S)$ & 4.2\,(1.8) &  4.2\,(1.8)  &   3.8\,(1.6) \\
		[1ex] 
		\hline 
	\end{tabular}
	\label{TabII} 
\end{table}

The calculations performed above can be compared to another theoretical approaches available in the literature. Let start comparing them to the STARlight Monte Carlo \cite{STARlight}. For the coherent production, the predictions for the $\Upsilon$ state ratios are lower that STARlight results. As discussed in Ref. \cite{GKMM}, the possible origin comes from the extrapolation of HERA-DATA and taking a fixed ratio for the distinct states in the Monte Carlo, whereas in current case the evolution on energy is dynamically generated by the parton saturation approach models and mostly by the meson wavefuntions for the radially excited states. The nuclear effects are also computed in a different way in the two formalisms. In STARLight, the nuclear shadowing is calculated using vector meson dominance (VDM) plus Glauber model for hadronic collisions. In our case, shadowing comes from the multiple scatterings of color dipoles and is described by the Glauber-Gribov approach. We verified that our results are also smaller that in Ref. \cite{STARlight}, which can be related to more shadowing in color dipole models compared to the VDM+Glauber approach.
In Ref. \cite{AN} only the coherent contribution was computed and the theoretical uncertainty we have found in the color dipole approach is comparable to  perturbative QCD formalism. Concerning similar dipole calculations, more recent investigations are available in Refs. \cite{GSSM2,NavarraUps}. In Ref. \cite{NavarraUps} only the coherent $\Upsilon (1S)$ production has been considered at 5.02 TeV. The results are smaller than ours and the main reason is the wavefunction chosen (Light Cone Gaussian wavefunction which gives smaller overall normalization compared to Boosted Gaussian one). The authors in \cite{NavarraUps} did not investigated the theoretical uncertainty associated to the wavefunction and dipole cross sections  as well (only the uncertainty coming from one model for dipole cross section was addressed). In Ref. \cite{GSSM2} the theoretical uncertainty for the coherent and incoherent cross section was investigated. However, predictions for higher energies in PbPb collisions were not presented and only the $\Upsilon (1S)$ state was considered (the results are consistent with ours in that case). Finally, we did not consider  photonuclear breakup in the present study. We will consider them in future analysis as they are important and the distinct channels have been measured for $\rho$ and $J/\psi$ photoproduction in UPCs \cite{RhoALICE,CMSJPSI}. This sort of analysis was recently done in Ref.\cite{Guzey}, where the coherent $\Upsilon (1S)$ production was considered using a pQCD model with NLO accuracy. An important point discussed in \cite{Guzey} is that the large $y$ region gives the dominant contribution for 0nXn and XnXn channel and they probe larger photon-target centre-of-mass energy than the case without neutron tagging.

\section{Summary}
 We presented the predictions of rapidity distribution and integrated cross sections for the  $\Upsilon(1S,2S,3S)$ states for the LHC run 2 energies. The rapidity intervals used in total cross section were selected to match with the rapidity coverage of LHCb and ALICE detectors both covering 2~$\leq y \leq 4.5$. The main contribution is the computation of the incoherent cross section within the color dipole approach and Glauber-Gribov treatment of nuclear shadowing. The cross section for the excited states are also calculated in a consistent formalism where the wavefunction of $2S$ and $3S$ states are theoretically well constrained. The usual procedure in the literature involves only an extrapolation of DESY-HERA production ratios to the LHC energies. Our calculations are directly comparable to the STARLight calculation, where distinct procedures are involved in the computation of nuclear shadowing (VDM plus Glauber model versus color dipole plus Glauber-Gribov approach) and how to obtain the incoherent cross section.

\begin{acknowledgments}
	This work was  partially financed by the Brazilian funding
	agency CNPq and Rio Grande do Sul funding agency FAPERGS.
\end{acknowledgments}


\begin{thebibliography}{99}

\bibitem{upcs} G. Baur, K. Hencken, D. Trautmann, S. Sadovsky, Y. Kharlov, Phys.
 	Rep. {\bf 364}, 359 (2002); C.~A. Bertulani, S.~R.~Klein and J.~Nystrand, Ann. Rev. Nucl. Part. Sci. {\bf 55}, 271 (2005).
 	
\bibitem{nik} N. N. Nikolaev, B. G. Zakharov,  Phys. Lett. B  {\bf 332}, 184 (1994); {Z. Phys. C} {\bf 64}, 631 (1994).
 	
\bibitem{Nemchik:1996pp} 
 	J.~Nemchik, N.~N.~Nikolaev, E.~Predazzi and B.~G.~Zakharov,
 	Phys.\ Lett.\ B {\bf 374}, 199 (1996).

\bibitem{CGC} 
  F.~Gelis, E.~Iancu, J.~Jalilian-Marian and R.~Venugopalan,
    Ann.\ Rev.\ Nucl.\ Part.\ Sci.\  {\bf 60}, 463 (2010);
  H.~Weigert,  Prog.\ Part.\ Nucl.\ Phys.\  {\bf 55}, 461 (2005); J.~Jalilian-Marian and Y.~V.~Kovchegov, Prog.\ Part.\ Nucl.\ Phys.\  {\bf 56}, 104 (2006); A.L. Ayala, M.B. Gay Ducati and E.M. Levin, Nucl.Phys. B {\bf 493}, 305 (1997).


\bibitem{heavy}  A. Dainese, in Proceedings of the 38th International Symposium of Multiparticle Dynamics (ISMD2008): Hamburg, Germany. September 15-20 2008, p.118-124, arXiv:0902.0377 [hep-ph]. 

 \bibitem{GKMM}	   M. B. Gay Ducati, F. K\"{o}pp, M.V. T. Machado, and S. Martins. Phys. Rev. D {\bf 94}, 094023 (2016).

 \bibitem{UpsilonLHCb} R. Aaij {et al.} [LHCb Collaboration], JHEP 1509, 084  (2015).    

\bibitem{Dusling}  K. Dusling , F. Gelis, T. Lappi and R. Venugopalan, Nucl. Phys.  A {\bf 836}, 159  (2010) .

\bibitem{Lappi} T. Lappi, H. M\"{a}ntysaari, R. Venugopalan, Phys. Rev. Lett. {\bf 114}, 082301 (2015). H. M\"{a}ntysaari and B. Schenke, Phys. Rev. Lett. {\bf117}, 052301 (2016). 

\bibitem{FlucHis}  L. Frankfurt, G.A. Miller and M. Strikman, 	Phys. Rev. Lett. {\bf 71}, 2859 (1993); A. Caldwell and H. Kowalski,  Phys.  Rev.  C {\bf 81}, 025203 (2010).

\bibitem{recworks} Kirill Tuchin, Phys. Rev.  C {\bf 79},  055206 (2009);   T. Lappi and H. M\"{a}ntysaari, Phys. Rev. C {\bf 83}
065202 (2011) ;  T. Toll and T. Ullrich, Phys.  Rev. C {\bf 87}, 024913   (2013) ;  H.  M\"{a}ntysaari, B. Schenke, C. Shen and P. Tribedy, Phys.  Lett.  B {\bf 772}, 681 (2017).
 
\bibitem{gbwnew} J. Cepila, J.G. Contreras, J. D. Tapia Takaki, Phys. Lett. {\bf B766 }, 186 (2017).

\bibitem{STARlight} S.R. Klein, J. Nystrand, J. Seger, Y. Gorbunov and J. Butterworth, Comput. Phys. Commun. {\bf 212}, 258 (2017).

\bibitem{KN} S. Klein and J. Nystrand, Phys. Rev. {\bf C60}, 014903 (1999).


\bibitem{Boris} B. Z. Kopeliovich and B. G. Zakharov, Phys. Rev. D 44, 3466 (1991); Yu.P.Ivanov, J.Huefner, B.Z.Kopeliovich and A.V.Tarasov, AIP Conf.Proc.  {\bf 660}, 283 (2003).

\bibitem{wfbg}
J.~Nemchik, N.~N.~Nikolaev, E.~Predazzi and B.~G.~Zakharov,
Z.\ Phys.\ C {\bf 75}, 71 (1997) 71.

\bibitem{Sanda2}  B.E. Cox, J.~R.~Forshaw and  R.~Sandapen, JHEP {\bf 0906 }, 034 (2009).


\bibitem{PRD59-014017} K. Golec-Biernat and M. W\"usthoff, Phys. Rev. {\bf D59}, 014017 (1998).

\bibitem{PLB590-199} E. Iancu, K. Itakura, S. Munier, Phys. Lett, {\bf B590}, 199 (2004).

\bibitem{PRD74-074016} H. Kowalski, L. Motyka and G. Watt, Phys. Rev. {\bf D74}, 074016 (2006).



\bibitem{Soyez} G. Soyez, Phys. Lett. {\bf B655 }, 32 (2007).

       		         			
\bibitem{Amir} A.H. Rezaeain and I. Schmidt, Phys. Rev. {\bf D88}, 074016   (2013).

\bibitem{GSSM2} G. Sampaio dos Santos and M.V.T. Machado, J. Phys.  {\bf G42}, 105001 (2015). 


\bibitem{armesto} N . Armesto,A. H. Rezaeian, Phys.Rev. {\bf D90},  054003 (2014).	

\bibitem{AN} A. Adeluyi and A. Nguyen, Phys. Rev. C {\bf 87}, 027901 (2013).            
			


\bibitem{NavarraUps} V.P. Gon\c{c}alves, B.D. Moreira and F.S. Navarra, Phys. Rev. {\bf D95}, 054011 (2017).		

  
\bibitem{RhoALICE} J. Adam {et al.} [ALICE Collaboration], JHEP 1509, 095  (2015).    	 

\bibitem{CMSJPSI} V. Khachatryan {et al.} [CMS Collaboration], Phys. Lett. B {\bf 396}, 772  (2017).

\bibitem{Guzey} V. Guzey, E. Kryshen and M. Zhalov, Phys. Rev. {\bf C93}, 055206 (2016).       			
             		
		 	  	
 
\end{thebibliography}
\end{document}